
\input harvmac
\def\inbar{\,\vrule height1.5ex width.4pt depth0pt}
\def\IB{\relax{\rm I\kern-.18em B}}
\def\IC{\relax\hbox{$\inbar\kern-.3em{\rm C}$}}
\def\ID{\relax{\rm I\kern-.18em D}}
\def\IE{\relax{\rm I\kern-.18em E}}
\def\IF{\relax{\rm I\kern-.18em F}}
\def\IG{\relax\hbox{$\inbar\kern-.3em{\rm G}$}}
\def\IH{\relax{\rm I\kern-.18em H}}
\def\II{\relax{\rm I\kern-.18em I}}
\def\IK{\relax{\rm I\kern-.18em K}}
\def\IL{\relax{\rm I\kern-.18em L}}
\def\IM{\relax{\rm I\kern-.18em M}}
\def\IN{\relax{\rm I\kern-.18em N}}
\def\IO{\relax\hbox{$\inbar\kern-.3em{\rm O}$}}
\def\IP{\relax{\rm I\kern-.18em P}}
\def\IQ{\relax\hbox{$\inbar\kern-.3em{\rm Q}$}}
\def\IR{\relax{\rm I\kern-.18em R}}
\font\cmss=cmss10 \font\cmsss=cmss10 at 7pt
\def\IZ{\relax\ifmmode\mathchoice
{\hbox{\cmss Z\kern-.4em Z}}{\hbox{\cmss Z\kern-.4em Z}}
{\lower.9pt\hbox{\cmsss Z\kern-.4em Z}}
{\lower1.2pt\hbox{\cmsss Z\kern-.4em Z}}\else{\cmss Z\kern-.4em Z}\fi}
\def\IGa{\relax\hbox{${\rm I}\kern-.18em\Gamma$}}
\def\IPi{\relax\hbox{${\rm I}\kern-.18em\Pi$}}
\def\ITh{\relax\hbox{$\inbar\kern-.3em\Theta$}}
\def\IOm{\relax\hbox{$\inbar\kern-3.00pt\Omega$}}

\font\cmss=cmss10

\def\ket#1{|{#1}\rangle}
\def\im{{\rm i}}

\Title{\vbox{\baselineskip12pt\hbox{UICHEP--TH/92--15;
FERMILAB--PUB--92/233--T; SLAC--PUB--5884}}}
{\vbox{\centerline{Novel Spin and Statistical Properties}
\centerline{of Nonabelian Vortices}}}

\centerline{Lee Brekke$^a$, Hans Dykstra$^b$, Adam F. Falk$^c$, and
Tom D. Imbo$^a$}

\footnote{}{$^a$Department of Physics, University of Illinois at Chicago,
Chicago IL 60607}

\footnote{}{$^b$ Fermi National Accelerator Laboratory, P.~O.~Box 500
Batavia IL 60510. Current Address: Department of Physics, University of
Massachusetts at Amherst, Amherst MA 01003}

\footnote{}{$^c$Stanford Linear Accelerator Center, Stanford CA 94309}

\vskip .7 in

\centerline{\bf Abstract}

\vskip 4pt

We study the statistics of vortices which appear in
$(2+1)$--dimensional spontaneously broken gauge theories, where a
compact group $G$ breaks to a finite nonabelian subgroup $H$.  Two
simple models are presented. In the first, a quantum state which is
symmetric under the interchange of a pair of indistinguishable vortices
can be transformed into an antisymmetric state after the passage
through the system of a third vortex with an appropriate $H$-flux
element. Further, there exist states containing two indistinguishable
{\it spinless} vortices which obey {\it Fermi statistics}. These
results generalize to loops of nonabelian cosmic string in $3+1$
dimensions. In the second model, fractional analogues of the above
behaviors occur. Also, composites of vortices in this theory may
possess fractional ``Cheshire spin'' which can be changed by passing an
additional vortex through the system.

\Date{}


In $2+1$ dimensions, it is well known that when a compact gauge group
$G$ is spontaneously broken via the Higgs mechanism to a finite
subgroup $H$, there are topologically stable vortices labeled by $h\in
H$ \ref\top{J.~Preskill, in {\it Architecture of the Fundamental
Interactions at Short Distances}, edited by P.~Ramond and R.~Stora
(North-Holland, Amsterdam, 1987).}.  When $H$ is nonabelian, there is a
long range interaction between these vortices due to a classical analog
of the (nonabelian) Aharonov--Bohm effect~\ref\ww {F.~Wilczek and
Y.-S.~Wu, Phys. Rev. Lett. {\bf 65} (1990) 13.}\ref\bu {M.~Bucher,
Nucl. Phys. {\bf B350} (1991) 163.}.  As a result, the classical and
quantum mechanics of such multi-vortex systems is richer and more
complex than that of their abelian counterparts. For example, the
concept of indistinguishability of vortices is quite subtle and leads
one to suspect that the statistical properties of these vortices may be
extremely unusual. In this paper we consider a model where, in fact,
the usual superselection rule between bosonic and fermionic statistics
is violated. That is, a quantum state symmetric under the interchange
of a pair of indistinguishable vortices can ``mix'' with an
antisymmetric state. (The existence of such behavior requires the
presence of one or more ``spectator'' vortices with appropriate
$H$-flux elements.) While recently it has been recognized that it is
possible to introduce such {\it ambistatistical\/} behavior into
certain identical particle systems~\ref\ambi{T.~D.~Imbo, C.~Shah Imbo
and E.~C.~G.~Sudarshan, Phys. Lett. {\bf B234} (1990) 103\semi
T.~D.~Imbo and J.~March-Russell, Phys. Lett. {\bf B252} (1990) 84.}, in
this theory it arises as a natural consequence of the long range vortex
interactions. The above model also possesses states containing two
indistinguishable spinless vortices obeying Fermi statistics, thus
violating the usual spin-statistics relation. For systems of three or
more indistinguishable vortices, parastatistical
states~\ambi\ref\prs{Y. Ohnuki and S. Kamefuchi, {\it Quantum Field
Theory and Parastatistics} (Springer-Verlag, Berlin, 1982)\semi
A.~P.~Balachandran, Syracuse University preprint SU-4428-454 (1990),
and references therein.} also can occur, and in the presence of a
spectator vortex different parastatistical types need not be
superselected. We proceed to construct a second model which displays
the fractional statistical analogue of each of the above behaviors. In
this theory, there are also violations of the superselection rule
between states with different values of $e^{2\pi\im J}$, where $J$ is
the total angular momentum. Our work may be considered as an
application and extension of the ideas in \bu , where a general
framework is developed for studying vortex statistics.

We now recall some basic facts concerning the physics of nonabelian
vortices.\footnote{$^1$}{In what follows, we work in temporal gauge
($A_0=0$).  Furthermore, we treat the vortices as pointlike objects and
never allow the positions of any pair of them to coincide.}  When
internal excitations are ignored, a single vortex configuration is
determined completely by the vortex position $x_1\in \IR^2$ and its
$H$-flux element $h$.  This flux element may be thought of as the
path-ordered exponential (or holonomy) of the corresponding gauge field
around a closed contour, based at a fixed point $x_0$ at spatial
infinity, which encircles the vortex once in a counterclockwise manner.
For two vortices at $x_1$ and $x_2$, there are two fluxes $h_1$ and
$h_2$ determined by path-ordered exponentials evaluated along contours
$C_1$ and $C_2$.  We take $C_i$ to encircle the vortex at $x_i$ as
shown in figure 1.  If the two vortices are moved along closed paths in
$\IR^2$, then when they return to their original positions their flux
elements in general will have changed \ww\bu .  In particular, if the
vortex originally at $x_1$ encircles the one at $x_2$ once
counterclockwise, then both fluxes are conjugated by the product
$h_1h_2$. That is, in the final configuration the vortex at $x_i$ has
flux $(h_1h_2)h_i(h_1h_2)^{-1}$.  One way to see this is by
transforming to a singular gauge in which the gauge potential is zero
everywhere except on two strings beginning at $x_1$ and $x_2$ and
stretching out to spatial infinity in different directions.  As the
$h_1$ vortex is brought around the $h_2$ vortex we deform the two
strings so that they never intersect.\footnote{$^2$}{Note that in order
for this path to be of finite action, the endpoints at spatial infinity
of the two strings must remain fixed.} In the final configuration these
strings will have become entangled, and the above result may be
obtained by recalculating the holonomies along $C_1$ and $C_2$. Note
that these flux changes have a completely topological origin; at no
point during the evolution has any vortex traversed a region of nonzero
field. Similarly, if we exchange counterclockwise the positions of the
two vortices with initial fluxes $h_1$ and $h_2$, then we end up with
an $h_1h_2h_1^{-1}$ vortex at $x_1$ and an $h_1$ vortex at $x_2$. The
above statements generalize in a natural way to a system of $N$
vortices.  The total flux of an $N$-vortex system, defined as the
path-ordered exponential along a counterclockwise contour which
encircles all of the vortices, is conserved during the evolution of the
system although the fluxes of individual vortices will in general
become conjugated. If the vortices are located at the positions $x_i$,
$1\leq i\leq N$, with fluxes $h_i$ (defined as in figure 1), then this
total flux is $h_1h_2\cdots h_N$.

There is a subtle point concerning the distinguishability of vortices.
Although two single-vortex fields with conjugate flux elements are
distinct configurations in the broken gauge theory, they cannot be
distinguished by any physical measurement. The standard procedure for
probing the flux element of a vortex is by an Aharonov-Bohm scattering
process involving particles carrying nontrivial $H$-charge. More
precisely, suppose we take such a particle once around a vortex of flux
$h$. If we assume that the particle is initially in a state $\ket{u}$
which transforms according to a faithful representation $D$ of $H$,
then upon its return it will be in the state $D(h)\ket{u}$. We may then
measure the matrix elements $\langle v|D(h)\ket{u}$ by interfering this
particle with a variety of other particles in states $\ket{v}$ also
transforming under $D$. In principle, it seems as though we may in this
way determine all matrix elements of $D(h)$, and hence $h$ itself.
However it has been shown that only the character
$\chi^{(D)}(h)=\sum_v\langle v|D(h)\ket{v}$ can actually be obtained in
this fashion, if one takes into account the effects of the creation and
annihilation of virtual vortex-antivortex pairs \ref\mon{M.~G.~Alford,
K.-M.~Lee, J.~March-Russell and J.~Preskill, Princeton University
preprint PUPT--91--1288 (1991).}. Since
$\chi^{(D)}(h_1)=\chi^{(D)}(h_2)$ if $h_1$ and $h_2$ are conjugate, we
see that vortices with conjugate flux elements are indistinguishable.
We note that this result holds even at energies well below the
threshold for the creation of a real vortex-antivortex pair.

{}From the above discussion it is clear that vortices possessing
conjugate flux elements should be considered as different internal
states of a single species $\alpha$. One consequence of this for the
quantum mechanics of such objects is that an isolated pair of vortices
of type $\alpha$ need not be bosons. Indeed, they can be anyons. We
will also see that in the presence of a third (distinguishable) vortex
whose flux does not commute with a representative flux from the
$\alpha$ species, the statistical behavior can be even more bizarre. In
general, a quantum mechanical $N$-vortex state containing $n_i$
vortices of type $\alpha_i$, $1\leq i\leq k$, is a linear combination
of states in which the vortices have well-defined fluxes. (We assume
that the vortices are localized about the positions $x_1,\dots x_N$ as
in figure 1, the first $n_1$ of these positions being associated with
species $\alpha_1$, the next $n_2$ with species $\alpha_2$, etc.) These
states may be obtained from each other by moving the vortices around
closed loops in the plane and/or permuting the positions of vortices of
the same species. We will denote by $\ket{h_1,\dots
,h_{n_1};h_{n_1+1},\dots , h_{n_1+n_2};\dots h_N}$ the term in this sum
in which the vortex localized about $x_i$ has flux $h_i$. Note that if
we let $N_i=\sum_{j=1}^{i}n_j$, then the fluxes $h_{N_i+1}$ through
$h_{N_{i+1}}$, for any fixed $0\leq i\leq k$ (where $N_0\equiv 0$),
must all lie in the same conjugacy class.

The loop and permutation operations in the above $N$-vortex system form
a group known as the ``partially colored braid group'' $B_{n_1,\dots ,
n_k}$ \ref\falk{L.~Brekke, A.~F.~Falk, S.~J.~Hughes and T.~D.~Imbo,
Phys. Lett. {\bf B271} (1991) 73.}. It can be interpreted as the
fundamental group of the configuration space of our effective
$N$-vortex system. The Hilbert space spanned by all possible states
$\ket{h_1,\dots ,h_{n_1};h_{n_1+1},\dots , h_{n_1+n_2};\dots h_N}$
decomposes into a direct sum of subspaces, each of which transforms
according to an irreducible unitary representation (IUR) of
$B_{n_1,\dots ,n_k}$. In each of these sectors, the statistics of the
vortices of type $\alpha_i$ is determined by the action of the
$\alpha_i$ exchange operations on the states.  The group $B_{1,1}$,
which is relevant for a system consisting of two distinguishable
vortices, is isomorphic to the integers $\IZ$, whose generator $\ell$
is taken to be the operation of circling the vortex at $x_1$
counterclockwise around the one at $x_2$. The action of $\ell$ on the
corresponding two-vortex states is given by
$\ell\ket{h_1;h_2}=\ket{gh_1g^{-1};gh_2g^{-1}}$, where $g=h_1h_2$. The
group $B_2$ which arises in a system of two indistinguishable vortices
is again isomorphic to $\IZ$, but this time the generator $\sigma$ is
the operation which exchanges the two vortices in a counterclockwise
manner. It acts as $\sigma\ket{h_1,h_2}=\ket{h_1h_2h_1^{-1},h_1}$. As a
final example, the group $B_{1,2}$ is an infinite discrete group which
can be generated by two elements $\sigma$ and $\ell$, representatives
of which are shown in figure 2, subject to the single defining relation
$(\sigma\ell)^2=(\ell\sigma)^2$. The action of these generators on the
three-vortex states is given by
$\ell\ket{h_1;h_2,h_3}=\ket{gh_1g^{-1};gh_2g^{-1},h_3}$ and
$\sigma\ket{h_1;h_2,h_3}=\ket{h_1;h_2h_3h_2^{-1},h_2}$, where
$g=h_1h_2$. Since this group is nonabelian it has IUR's of dimension
greater than one, which will lead to exotic statistical behavior in the
associated Hilbert space sectors.

We will now illustrate these features of multi-vortex systems with two
simple examples. In the first model the group $G=SU(2)$ is broken down
to the eight element quaternion group $H=Q_8$ \ref\pk{J.~Preskill and
L.~M.~Krauss, Nucl. Phys. {\bf B341} (1990) 50.}. The elements of
$Q_8\subset SU(2)$ are $\pm 1$ and the Pauli matrices
$a_j=\pm\im\tau_j$, $j=1,2,3$. There are five conjugacy classes, given
by $\{ 1\}$, $\{ -1\}$ and $\{\pm a_j\}$. We start with the state
$\ket{a_3,-a_3}$ containing two vortices of a single species $\beta$.
The total flux of the system is trivial. A basis for a representation
of the braid group $B_2$ can be obtained from the states
$\sigma^m\ket{a_3,-a_3}$ for integer $m$. Even powers of $\sigma$
operate trivially on this state, while odd powers yield
$\ket{-a_3,a_3}$.\footnote{$^3$}{By allowing the action of $\sigma$ on
$\ket{a_3,-a_3}$ to include a phase, it might seem that
$\ket{a_3,-a_3}$ and $\sigma^2\ket{a_3,-a_3}$ need only be equal up to
a phase. However, here $\sigma^2$ (unlike $\sigma$) can be thought of
as a loop in the {\it full} gauge theory configuration space. The phase
difference between the above two states is then just a ``Berry's
phase'' with respect to this loop. But the gauge theory configuration
space does not possess the appropriate topology for the existence of
such nontrivial phases. Thus the above phase is 1. Similar comments
hold for the other examples in this paper.} This two-dimensional
representation of $B_2$ can be decomposed into a direct sum of two
one-dimensional representations by forming the symmetric and
antisymmetric combinations of the above states. In the corresponding
Hilbert space sectors, the $\beta$ vortices are bosons and fermions
respectively. The appearance of Fermi statistics is somewhat surprising
since all of the fundamental fields in our theory are bosonic. (We will
return to this point later.) If we start with a state of three or more
$\beta$ vortices, not all of the same flux, then a similar procedure
will also yield parastatistical Hilbert space sectors.

Next, consider the three-vortex state $\ket{a_1;a_3,a_3}$ in which a
single type $\alpha$ vortex (at $x_1$) has flux $a_1$, and two type
$\beta$ vortices (at $x_2$ and $x_3$) have flux $a_3$. The total flux
of the system is $-a_1$. We obtain a basis for a representation of the
group $B_{1,2}$ by operating on this state with all combinations of the
generators $\sigma$ and $\ell$ (and their inverses).  This basis is
\eqn\basis{\eqalign{
    &\ket{a_1;a_3,a_3},\cr
    &\ket{-a_1;-a_3,a_3}=\ell\ket{a_1;a_3,a_3},\cr
    &\ket{-a_1;a_3,-a_3}=\sigma\ell\ket{a_1;a_3,a_3},\cr
    &\ket{a_1;-a_3,-a_3}=\ell\sigma\ell\ket{a_1;a_3,a_3}.\cr}}
Operating further with $\sigma$ or $\ell$ simply permutes these four
states. This four-dimensional representation of $B_{1,2}$ reduces to a
direct sum of two one-dimensional IUR's and one two-dimensional IUR.
This can be seen in the transformed basis
\eqn\trans{\eqalign{
    &\ket\psi={1\over2}\big(\ket{a_1;a_3,a_3}+\ket{a_1;-a_3,-a_3}
     +\ket{-a_1;-a_3,a_3}+\ket{-a_1;a_3,-a_3}\big),\cr
    &\ket\xi={1\over2}\big(\ket{a_1;a_3,a_3}+\ket{a_1;-a_3,-a_3}
     -\ket{-a_1;-a_3,a_3}-\ket{-a_1;a_3,-a_3}\big),\cr
    &\ket{\phi_1}={1\over\sqrt{2}}
     \big(\ket{a_1;a_3,a_3}-\ket{a_1;-a_3,-a_3}\big),\cr
    &\ket{\phi_2}={1\over\sqrt{2}}
     \big(\ket{-a_1;-a_3,a_3}-\ket{-a_1;a_3,-a_3}\big).\cr}}
The vectors $\ket\psi$ and $\ket\xi$ carry one-dimensional IUR's
$\rho_1$ and $\rho_2$ respectively, while $\ket{\phi_1}$ and
$\ket{\phi_2}$ form a basis for a two-dimensional IUR $\rho_3$. The
generators $\sigma$ and $\ell$ in these representations are
\eqn\matr{\eqalign{
    &\rho_1(\sigma)=1,\cr&\rho_2(\sigma)=1,\cr&\rho_3(\sigma)=
    \pmatrix{1&0\cr0&-1},\cr}\qquad
    \eqalign{&\rho_1(\ell)=1,\cr&\rho_2(\ell)=-1,\cr&\rho_3(\ell)=
    \pmatrix{0&1\cr1&0}.\cr}}
In the sectors corresponding to $\ket\psi$ and $\ket\xi$, the vortices
of type $\beta$ are bosons, since $\rho_1(\sigma)=\rho_2(\sigma)=1$. In
the $\rho_3$ sector they obey {\it ambistatistics}, which is neither
Bose nor Fermi but contains aspects of each \ambi .

It is interesting to see what happens to the states $\ket{h_1;h_2,h_3}$
when we remove one of the vortices from the system.  First, we note
that the resulting two-vortex state will depend on the manner in which
the third vortex is removed. That is, there is a variety of ways in
which this vortex may be threaded through the others as it is dragged
off to spatial infinity, and the total flux of the remaining two-vortex
system will depend on the choice of this path \bu .  If we choose to
drag the vortex at $x_3$ in figure 2 off to the right, then
$\ket{h_1;h_2,h_3}$ will be reduced to the two-vortex state
$\ket{h_1;h_2}$.  Applying this procedure to the states in \trans, we
find
\eqn\twovor{\eqalign{
   &\ket\psi \to {1\over\sqrt{2}} \big( \ket{\psi_+} + \ket{\psi_-}
    \big)={1\over2}\big( \ket{a_1;a_3} + \ket{-a_1;-a_3} \big)
    +{1\over2}\big( \ket{a_1;-a_3} + \ket{-a_1;a_3} \big),\cr
   &\ket\xi \to {1\over\sqrt{2}} \big( \ket{\xi_+} + \ket{\xi_-}
    \big)={1\over2} \big( \ket{a_1;a_3} - \ket{-a_1;-a_3} \big)
    +{1\over2} \big( \ket{a_1;-a_3} -\ket{-a_1;a_3} \big),\cr
   &\ket{\phi_1} \to {1\over 2}\big (\ket{\psi_+}-\ket{\psi_-}
    +\ket{\xi_+}-\ket{\xi_-}\big ),\cr
   &\ket{\phi_2} \to {1\over 2}\big (\ket{\psi_+}-\ket{\psi_-}
    -\ket{\xi_+}+\ket{\xi_-}\big ).\cr}}
Note that the two-vortex states $\ket{\psi_\pm}$ and $\ket{\xi_\pm}$
have total flux $\pm a_3a_1$. Each of these four states defines a
one-dimensional IUR of the two-vortex braid group $B_{1,1}=\IZ$
generated by the single loop $\ell$ in figure 2; $\ell$ operates
trivially on $\ket{\psi_\pm}$ and multiplies $\ket{\xi_\pm}$ by $-1$.
Thus the total Hilbert space breaks up into a direct sum of four
pieces, each of which corresponds to one of these representations and
has a fixed total flux.  In fact, all two-vortex states with a type
$\alpha$ vortex at $x_1$ and a type $\beta$ vortex at $x_2$ appear
here. Similarly, we can reduce the three-vortex states in \trans\ by
removing the $\alpha$ vortex at $x_1$ off to the left, in which case we
again obtain a basis for all two-vortex states (both being of type
$\beta$ now). Note that $\ket\psi$, $\ket\xi$ and $\ket{\phi_1}$ yield
bosonic two-vortex states, while $\ket{\phi_2}$ gives a fermionic
state. In turn, it is straightforward to verify that all three-vortex
states (including the ambistatistical ones in \trans ) may be obtained
by removing vortices from a larger system. In such a system, other
superselection rules may be violated, such as that between Bose
statistics and parastatistics, or between different forms of
parastatistics.

The above analysis can be used to shed light on the nature of
ambistatistics in this model. Consider an $a_1$ vortex which is
incident from the left on the symmetric two-vortex state $\big
(\ket{a_3,a_3} - \ket{-a_3,-a_3}\big )$, and eventually combines with
it to make the total state $\ket{\phi_1}$.  If the $a_1$ vortex then
loops around one (say the leftmost) of the other vortices, then
$\ket{\phi_1}$ will become $\ket{\phi_2}$. So if we now remove the
$a_1$ vortex as it came, the two $a_3$ vortices will be left in the
anti-symmetric state $\big (\ket{-a_3,a_3} - \ket{a_3,-a_3}\big )$.
More generally, in the quantum mechanical scattering of an $a_1$ vortex
off of the above bosonic state, this path along with many others will
contribute. After the incident vortex has been scattered off to
infinity, we will be left with a nonzero probability that the two
remaining indistinguishable vortices are fermions. In this sense,
ambistatistics violates the usual superselection rule between bosons
and fermions.  We stress that here this behavior is a natural
consequence of the physics of nonabelian vortices, and has not been
introduced in an {\it ad hoc\/} manner.

We now turn to a model where the group $G=SU(3)$ is broken down to the
six-element group $S_3$, the permutation group on three objects, which
is generated by the matrices
\eqn\mat{t_1=\pmatrix{0&1&0\cr 1&0&0\cr 0&0&-1},\ \ \
    t_2=\pmatrix{-1&0&0\cr0&0&1\cr 0&1&0}.}
$S_3$ has three conjugacy classes, given by $\{ 1\}$, $\{ t_1,\ t_2,\
t_1t_2t_1 \}$ and $\{ t_1t_2,\ t_2t_1\}$. We start with the state
$\ket{t_1,t_2}$, which contains two indistinguishable vortices since
the fluxes $t_1$ and $t_2$ are conjugate. The total flux of the system
is $t_1t_2$. We obtain a basis for a representation of the braid group
$B_2$ by operating on this state with all powers of the vortex exchange
operator $\sigma$. This basis is
\eqn\any{\eqalign{&\ket{t_1,t_2},\cr &\ket{t_1t_2t_1,t_1}
     =\sigma\ket{t_1,t_2},\cr
    &\ket{t_2,t_1t_2t_1}=\sigma^2\ket{t_1,t_2}.\cr }}
Operating once more with $\sigma$ brings us back to the original state.
This three-dimensional representation of $B_2$ decomposes into a direct
sum of three one-dimensional representations. This can be seen by
transforming to the basis
\eqn\ab{\eqalign{&\ket{\lambda_0}={1\over\sqrt{3}}\big (\ket{t_1,t_2}+
     \ket{t_1t_2t_1,t_1}+\ket{t_2,t_1t_2t_1}\big ),\cr
    &\ket{\lambda_1}={1\over\sqrt{3}}\big (\ket{t_1,t_2}+
     \omega^2\ket{t_1t_2t_1,t_1}+\omega\ket{t_2,t_1t_2t_1}\big ),\cr
    &\ket{\lambda_2}={1\over\sqrt{3}}\big (\ket{t_1,t_2}+\omega
     \ket{t_1t_2t_1,t_1}+\omega^2\ket{t_2,t_1t_2t_1}\big ),\cr }}
where $\omega =\exp (2\pi\im /3)$. Each of the above states carries a
one-dimensional representation of $B_2$ given by
$\sigma\ket{\lambda_m}=\omega^m\ket{\lambda_m}$, $m=0,1,2$. In the
Hilbert space sector spanned by $\ket{\lambda_0}$ the two
indistinguishable vortices are bosons. In the $\ket{\lambda_1}$ and
$\ket{\lambda_2}$ sectors they obey {\it fractional statistics} with
statistical angles $\theta_1 = 2\pi /3$ and $\theta_2=4\pi /3$
respectively. Once again, this is somewhat surprising since all of the
fundamental fields in our theory are bosonic. If we add one or more
additional vortices from the $t_1$ conjugacy class to the above system,
we will also obtain parastatistical sectors as well as their fractional
generalizations (nonabelian anyons).

To help us understand these results, it is useful to introduce the
notion of Cheshire charge. In a vortex system with total flux $h$, the
only globally well-defined gauge transformations are those belonging to
the subgroup $C_H(h)\subseteq H$ consisting of all elements that
commute with $h$. States of the system may carry a nontrivial
representation of $C_H(h)$, although this ``Cheshire'' charge is a
global property of these states and is not carried by any individual
vortex \pk\ref\prop{M.~G.~Alford, S.~Coleman and J.~March-Russell,
Nucl. Phys. {\bf B351} (1991) 735.}. Since the gauge transformations in
$C_H(h)$ commute with elements of the appropriate braid group, we can
choose states which simultaneously carry a representation of both
groups. For the states \ab\ in the $S_3$ model, the globally
well-defined subgroup $C_{S_3}(t_1t_2)$ is generated by $t_1t_2$ and is
isomorphic to the cyclic group $\IZ_3$. The state $\ket{\lambda_0}$
transforms trivially under this $\IZ_3$, while $\ket{\lambda_1}$ and
$\ket{\lambda_2}$ pick up the phases $\omega^2$ and $\omega$
respectively under a gauge transformation by $t_1t_2$. Thus, these
anyonic states carry a nontrivial $\IZ_3$ charge. The existence of this
charge can shed some light on the emergence of fractional statistics in
this example. Viewed from far away, these two-vortex systems look like
composites of a point vortex with flux $t_1t_2$ and a nontrivial
$\IZ_3$ charge. Rotating this composite by $2\pi$ (counterclockwise)
has the same effect as performing a $t_1t_2$ gauge transformation. Thus
the state $\ket{\lambda_1}$ (for instance), which picks up the phase
$\omega^2\equiv e^{2\pi\im J}$, has total angular momentum $J=2/3$ +
integer. The fractional part of $J$ may be called ``Cheshire spin''
since it is a direct consequence of the Cheshire charge in the
composite system and inherits similar nonlocal properties. From the
point of view of the two original indistinguishable vortices which are
contained in this ``charged flux tube'', there are three contributions
to $J$ \falk . The first two are the individual spins of the
constituent vortices themselves. Since they are pure flux tubes, these
contributions vanish. Finally, there is a contribution from one vortex
encircling the other counterclockwise. This must give the entire spin
of the system. However, this is the same as performing the double
exchange operation $\sigma^2$ on $\ket{\lambda_1}$. Thus, a single
exchange of the two constituent vortices must yield a phase $\pm
e^{\pi\im J}$. Our earlier analysis shows that the plus sign is
chosen. Note that the constituent vortices are spinless particles
obeying fractional statistics, that is, they violate the usual
spin-statistics relation.\footnote{$^4$}{Cheshire charge is also
responsible for the emergence of Fermi statistics in the $Q_8$ model.
Moreover, for the states in \trans\ the globally well-defined subgroup
$C_{Q_8}(-a_1)$ is generated by $a_1$ and is isomorphic to the cyclic
group $\IZ_4$. The states $\ket{\psi}$ and $\ket{\xi}$ transform
trivially under this $\IZ_4$, while $\ket{\phi_1}$ and $\ket{\phi_2}$
change sign under a gauge transformation by $a_1$. Thus the
ambistatistical sector of the Hilbert space carries a nontrivial
$\IZ_4$ Cheshire charge.} This result can be traced directly to the
nonabelian nature of the vortices and the corresponding nonlocal
interactions between them.

Systems with Cheshire behavior may violate the superselection rule
between states with different statistical angles, as well as between
states with different values of $e^{2\pi\im J}$ . For example,
consider the passage of a vortex of flux $t_1t_2$ between the two
vortices in the state $\ket{\lambda_1}$. It is straightforward to show
that in the resulting state we have $\sigma =1$ (the two constituent
vortices are now bosons) and $e^{2\pi\im J}=1$ (the composite now has
integral angular momentum). The missing fractional spin is carried away
as orbital angular momentum by the incident vortex as it moves off to
infinity. In general, the passage of a vortex with a noncommuting flux
from a distinct species can also change a {\it nonabelian} anyonic
state into a state with different statistical properties.

Throughout this paper we have considered systems in which the number of
vortices is constant, and we have neglected the creation of real
vortex-antivortex pairs.  It is reasonable to ask whether such
processes impose constraints on the allowed multi-vortex states, and
thereby possibly rule out certain types of exotic statistical behavior.
Indeed, recent studies have indicated that only Bose, Fermi and
fractional statistics are allowed for identical particles in the
presence of pair creation, and furthermore that the usual
spin-statistics relation holds for such particles
\ref\spin{A.~P.~Balachandran, A.~Daughton, Z.-C.~Gu, G.~Marmo,
R.~D.~Sorkin and A.~M.~Srivastava, Mod. Phys. Lett. {\bf A5} (1990)
1575\semi R.~D.~Tscheuschner, Int.~J.~Theor. Phys. {\bf 28} (1989)
1269.}. However this work assumes that the particles carry no internal
label. Hence it does not apply to parastatistical systems, for example,
where it is known that consistent relativistic quantum field theories
can be formulated \prs . The nonabelian vortex systems discussed here
also avoid such restrictions, since two vortices of the same species
can have distinct fluxes.  Thus we believe that our results concerning
the exotic statistics of vortices are not an artifact of our quantum
{\it mechanical\/} approach.

We close with two brief comments.  First, an analysis similar to that
presented here may be applied to systems with other symmetry breaking
schemes $G\to H$. While such a treatment generally will be more
cumbersome, the emergence of exotic statistics for indistinguishable
nonabelian vortices is generic.  More precisely, an isolated system of
indistinguishable vortices having noncommuting flux elements generally
will possess anyonic states (possibly even {\it nonabelian} anyonic
states if there are more than two vortices). States containing
indistinguishable vortices in the presence of a spectator vortex with a
noncommuting flux generally will violate the superselection rules
between the above statistical behaviors, as in ambistatistical systems.
Which representations of the appropriate braid group $B_{n_1,\dots
,n_k}$ are actually realized depends on the symmetry breaking scheme.
Second, it is interesting to note that nonfractional exotic behaviors
such as ambistatistics and parastatistics are in no way peculiar to
$2+1$ dimensions. Similar results hold for closed loops of nonabelian
cosmic string in the analogous $(3+1)$-dimensional spontaneously broken
gauge theories. Here, the analogue of bringing one vortex around
another is threading one loop of cosmic string through another loop.

We would like to thank John March-Russell for helpful comments and
interesting discussions. This work was supported in part by DOE
contracts DE-FG02-84ER40173 (L.~B. and T.~I.) and DE-AC03-76SF00515
(A.~F.). Most of this work was done while T.~I. was a Junior Fellow in
the Harvard Society of Fellows.

\listrefs

\bye